# Quantum nonlocality and biological coherence

S.N. Mayburov

Lebedev institute of physics, Leninski pr., 53, Moscow, Russia, mayburov@sci.lebedev.ru

## Abstract

Quantum-mechanical nonlocality considered as possible mechanism of long-distance correlations in living organisms and plants, which regulate their coherent development and functioning. It's shown that Doebner-Goldin nonlinear quantum formalism permits to extend nonlocality effects beyond standard EPR-Bohm scheme and due to it supposedly can describe long-distance correlations in some biological processes. Comparison of model calculations with some experimental results discussed.

## Introduction

It's well known that development and functioning of large biological systems performed quite consistently even at large distances between their parts. For example, kidney and liver cells, blood erythrocytes identify and attract the proper partner cells and reject wrong ones at the distance of several microns, which are much larger than the range of chemical forces (Vitiello, 2001). Another notorious example is morphogenesis problem, i.e. proportional and optimal growth of multicell plants and organisms. Up to now mechanism which regulates spontaneous cell division in necessary and optimal way at significant distances between them is poorly understood. It's difficult to admit that such long-distance coherence can be achieved via chemical messengers only, so it was argued long ago that some other physical mechanism can be responsible for that. In particular, it was proposed that coherent electromagnetic field, similar to laser field, can exist inside biosystem and perform such communications (Frohlich, 1968; Popp, 1998). However, it seems doubtful that such electromagnetic field coherence can be conserved in wide frequency range during long time periods of the order of minutes and hours. To support biological coherence paradigm, Frohlich and others applied the methods of solid state physics and quantum optics to biological systems, but their arguments don't seem watertight (Frohlich, 1968; Vitiello, 2001). In principle, such functions can be performed also by intricate noncoherent electromagnetic signaling mechanism between organism constituents, but no conclusive theory for that was proposed (Kucera and Cifra, 2013).

Meanwhile, alternative idea explaining biological coherence origin was proposed earlier, namely, that it induced by fundamental nonlocality of quantum mechanics (QM) (Primas, 1982; Josephson, 1991; Bischof, 2003). This phenomenon was first formulated as famous EPR-Bohm paradox and later developed as Bell theorem (Jauch, 1968; Peres, 2002; Bell, 2004). Now these effects confirmed experimentally and applied in quantum communications and computing (Sudbery, 1986; Peres, 2002). In fact, this mechanism permits to realize some form of nonlocal correlations (NC) or 'action at the distance' between distant subsystems of quantum system $S$. NC realization in EPR-Bohm variant demands that initially $S$ subsystems $S_1, ..., S_n$ should interact with each other, after such interaction seized, such NC conserve the correlations between $S_i$ uncertain parameters even at large distance between those $S$ subsystems. Obviously, such conditions is quite difficult to fulfill for communications in dense and warm biological systems. Therefore, particular EPR-Bohm nonlocality mechanism doesn't look suitable candidate for biological coherence explanation. However, it was supposed for long that quantum NC can be more general concept than standard QM formalism admits and, in principle, some other NC effects, besides EPR-Bohm one, can exist (Stapp, 1997; Cramer, 1986; Bell, 2004; Norsen, 2009; Korotaev, 2011). Below we'll discuss experimental results which presumably support this hypothesis, basing on them novel NC mechanism is proposed (Mayburov, 2021b; Mayburov, 2022). Note that this approach is principally different from hidden parameter theories (Sudbery, 1986).

Let's consider first the conditions to which such nonstandard NC mechanism should obey in general. Plainly, beside causality demands, such NC should agree with all standard invariance principles i.e. time, space shift and rotation symmetries. Analogously to EPR-Bohm NC effect, we'll suppose that such NC by itself can't transfer the energy, momentum or orbital momentum between distant objects. Hence the system average energy, momentum and orbital momentum should not change during such communications. It will be shown below that application of nonlinear NC Hamiltonians permit to obey to such constraints (Doebner and Goldin, 1992; Mayburov, 2021a, Mayburov, 2022) . More complicated NC model variants for which these constraints are absent will be considered elsewhere. Let's analyze which quantum systems can be most vulnerable to such NC influence. Suppose that states of two distant evolving objects $S_1$, $S_2$ during arbitrary time interval $\{t_0, t_f\}$ become to differ from standard QM predictions due to their conjugate NC influence. From the described assumptions it follows that according to QM rules the initial $S_1$, $S_2$ states can't be stationary, because such states are ground ones and possess the minimal energy, so only some energy transfer can make them to evolve to another state which will be excited one. Hence the reasonable option is that $S_1$, $S_2$ are degenerate systems, i.e. they have several states with the same energy and during interval $\{t_0, t_f\}$ can evolve from initial state to another one with the same energy. In this case, due to NC influence some $S_1$, $S_2$ parameter values, in principle, can deviate from standard QM predictions. The simple example of such system is the particle with energy level $E$ confined in symmetric double well potential divided by potential wall $U_0$ with $E \leq U_0$. Suppose that such $S_1$ has two degenerate states $g_1$, $g_2$ in these two wells with the same energy $E$, and at $t_0$ it's in state $g_1$ confined in one well. Thereon, due to under-barrier tunneling it would spread gradually into other well, so that it will evolve with the time to some $g_1$, $g_2$ superposition (Peres, 2002). In this case, the hypothetic $S_2$ NC influence on $S_1$ can change its tunneling rate in comparison with QM predictions, so that it would influence $S_1$ state parameters, in particular, resulting $g_1$, $g_2$ probabilities at $t_f$. Due to reciprocal $S_1$ NC influence, analogous variations would occur with $S_2$ if it possesses similar degenerate structure. In this paper, phenomenological model of such NC processes involving quantum tunneling will be considered, in its framework, NC effects described by QM evolution equations with specific nonlinear Hamiltonians (Mayburov, 2021b, Mayburov, 2022).

### Experimental motivations

Experiments considered here concerned with situations when quantum system suffers abrupt transformations, in particular, it occurs in chemical and nuclear reactions, for example, decays of unstable nuclei. It was acknowledged previously that due to strong nuclear forces no environment influence can change the isotope decay parameters significantly (Martin, 2011). However, recent results indicate that some cosmophysical factors related to Earth motion along its orbit can influence them, in particular, their life-time and decay rate at $10^{-3} – 10^{-4}$ scale (Alekseev, 2016; Fischbach, 2009). First results, indicating the deviations from exponential β-decay rate dependence, were obtained during the precise measurement of $^{32}$Si isotope life-time . In addition to standard decay exponent, sinusoidal annual oscillations with the amplitude about $6*10^{-4}$ relative to total decay rate and maxima in the end of February, were found during 5 years of measurements. Since then, annual oscillations of β -decay rate for different heavy nuclei from Ba to Ra were reported, for most of them the oscillation amplitudes are of the order $5*10^{-4}$ with their maximum on the average at mid-February (Fischbach, 2009). Life-time of α -decay isotopes $^{212}$Po, $^{213}$Po, $^{214}$Po was measured directly, the annual and daily oscillations with amplitude of the order $7*10^{-4}$ and annual minima at March for different isotopes and daily maxima around 6 a.m. for $^{214}$Po were found during 6 years of measurements (Alekseev, 2016). It was shown also that decay rates of $^{53}$Mn, $^{55}$Fe *e*-capture and $^{60}$Co *β*-decay correlate with solar activity, in particular, with intense solar flare moments, preceding them for several days; in this case, observed decay rate variations are of the order $10^{-3}$ (Fischbach, 2009; Bogachev 2020).

Parameters of some chemical reactions also demonstrate the similar dependence on solar activity and periodic cosmophysical effects. First results were obtained for bismuth chloride hydrolysis, its reaction rate was shown to correlate with solar Wolf number and intense solar flare moments (Picccardi, 1962). It was demonstrated that for biochemical unithiol oxidation reaction its rate correlates with solar activity, in particular, with intense solar flares and it grows proportionally to Wolf number. It was found also that its rate correlates with periodic Moon motion and Earth axis nutation (Troshichev, 2004). Takata biochemical blood tests also indicate strong influence of solar activity and Sun position in the sky (Takata, 1951). It performed via human blood reaction with sodium carbonite $Na_2CO_3$ resulting in blood flocculation. Its reaction rate parameter demonstrates fast gain for blood samples taken from organism starting 6-8 minutes before astronomic sunrise moment, such gain continues during next 30-40 minutes. Such reaction parameter behavior is independent of mountain or cloud presence at eastern horizon sector, which can screen the Sun. This parameter demonstrates approximate invariance for blood taken during the solar day and slow decline after sunset, such daily dependence conserved even in complete isolation from electromagnetic fields and solar radiation. This parameter also demonstrates the gain for larger solar Wolfe numbers and for the test location shift in the direction of Earth equator. Significant parameter correlation with Sun eclipse moments also was observed; some of these results were confirmed by other groups (Gierhake, 1938; Jezler, 1938; Koller, 1938; Kaulbersz et al,1958; Sarre, 1951). Taking blood samples at different altitudes up to several km, Takata concluded that the influence source isn't the Sun itself, but Earth atmosphere at the altitude above 6-7 km (Kiepenheuer, 1950; Takata, 1951). It's established now that during solar day at such altitudes the intense photochemical reactions occur, in particular, $O_2$, $SO_2$, $NO_2$ molecule destruction by UV solar radiation, which results in ozone and other compound synthesis (McEven and Phillips, 1975). Hence it can be supposed that those chemical reactions induce NC influence which changes blood reaction activity in distant human organisms promptly. In supposedly explains the start of reaction rate gain 6-8 min. before astronomic sunrise and its independence of mountain or cloud presence. Really, at that time solar radiation already reach Earth atmosphere at such altitudes, whereas mountains and clouds can't absorb it being essentially lower. Similar daily variations were reported for deuterium diffusion into palladium crystal (Scholkman et al, 2012).

Experiments of other kind also exploit some biochemical and organic-chemical reactions, the example is reaction of ascorbic acid with dichlorphenolindophenol. Authors noticed first that dispersion of their reaction rates can change dramatically from day to day, sometimes by one order of value, whereas average reaction rate practically doesn't change (Shnoll, 1973; Shnoll and Kolombet, 1980; Shnoll, 2009) . Further studies have shown connection of this effect with some cosmophysical factors, first of all with solar activity, solar wind and orbital magnetic field. In particular, average rate dispersion becomes maximal during solar activity minima of 11 year solar cycle (Shnoll and Chetvericova, 1973; Shnoll, 2009). It supposes that high solar activity makes such chemical reaction process more ordered and regular in time. Shielding of chemical reactors from external electromagnetic field in iron and permalloy boxes practically don't change the reaction dispersions, hence such cosmophysical influence can't be transferred by electromagnetic fields. Individual nuclear decay or chemical reaction acts normally are independent of each other; such stochastic processes called Poissonian and described by Poisson probability distribution (Korn and Korn, 1968). For this distribution, at any time interval $dT$ the dispersion of decay count number $\sigma_p = N^{\frac{1}{2}}$, where $N$ is average count number per $dT$. If the resulting dispersion $\sigma < \sigma_p$, it means that this process becomes more regular and self-ordered and described by sub-Poisson statistics with corresponding distribution (Mandel and Wolf, 1995). For $\sigma \to 0$ the time intervals between events become fixed, i.e. process tends to become maximally ordered. If on the opposite $\sigma_p < \sigma$, such process corresponds to super-Poisson statistics, which is typical to collective chaotic processes. In both cases it can be supposed that solar activity acts on reaction volume as the whole, similarly to crystal lattice excitations. For quantum systems such dispersion variations are typical for squeezed states, below this approach will be applied in our model (Paul, 1982; Mandel and Wolfe, 1995). . In general, these considerations are applicable to arbitrary probability distributions of studied systems not only Poisson-like ones. Described analysis evidences that high solar activity makes molecular systems to perform chemical reactions in more self-ordered and regular way. Summing up, cited results permit to suppose the existence of unknown distant interactions of nonelectromagnetic origin, which supposedly can be attributed to hypothetical NC effects.

## Microscopic NC model

Experimental results considered here evidence that observed distant influence occur for evolving quantum systems when such system by itself suffers significant transformation even without hypothetical NC influence, in particular, it occurs for chemical reactions and decays of unstable nuclei. In standard QM formalism the system evolution operator defined mainly by quantum-to-classical correspondence (Jauch, 1968), for NC effects such guidelines are absent and we'll construct it basing only on general QM principles and cited experimental results. We'll construct here phenomenological model of NC effects, it can't describe directly discussed experimental results, but simulate analogous effects for simple quantum systems. Below we'll consider NC model for the system $\{A_i\}$ of $N$ unstable α-decay nuclei, its initial state is $\{A_i\}$ product state. In fact, the similar considerations are applicable to the evolution of arbitrary metastable systems, like atoms, molecules or biological systems, yet for nucleus α-decay its description is most simple. In QM formalism the system in pure state described by Dirac complex vector $\psi$ in Hilbert space denoted also as ket-vector $|\psi>$. For any system in pure state $\psi$ its evolution described by Schroedinger equation

$$i\hbar \frac{\partial \psi}{\partial t} = H\psi \qquad (1)$$

where $\hbar$ is Plank constant, $H$ is Hamiltonian operator (Peres, 2002). Gamow theory of nucleus α-decay admits that before the decay act occurs, α-particle with average kinetic energy $E$ already exists as independent entity inside nucleus (Gamow, 1928; Newton, 1961). It supposed that Coulomb and nucleus forces constitute potential barrier with maximal height $U_m$ on the nucleus periphery such that $E \leq U_m$. Hence α-particle can leave nucleus volume only by means of underbarrier tunneling, which explains large life-times of such isotopes. Therefore, alike for double well example, the state energy is the same inside and outside nucleus, and corresponding inside-outside states $\psi_{0,1}$ are degenerate and orthogonal to each other. Such energy degeneration permits, in principle, for some hypothetic NC mechanism to change nucleus decay parameters without any energy transfer to α-particle, but just changing the barrier transmission rate, such mechanism considered below. In Gamow theory, α-particle Hamiltonian

$$H = \frac{\vec{P}^2}{2m} + U(r) \qquad (2)$$

where $\vec{P} = -i\hbar\vec{\nabla}$ is its momentum operator, $m$ is α-particle mass, $U$ is nucleus barrier potential, $r$ is distance from nucleus center (Gamow, 1928; Newton, 1961). If at $t_0$ α-particle was in initial state $\psi_0^i = \psi_0$, inside $A_i$ nucleus, then the solution of eq. (1) for $A_i$ state $\psi^i(t)$ gives for decay probability rate at given moment $t$ for nucleus $A_i$

$$p_i(t) = \lambda \, exp[-\lambda \, (t-t_0)] \qquad (3)$$

which is time derivative of total $A_i$ decay probability from $t_0$ to $t$, here $\lambda$ depends on $m$, $U$ and on Gamow kinematic factor $n_d$ related to α-particle motion inside nucleus (Gamow,1928; Newton, 1961). Thus, at $t \to \infty$ nucleus state evolves to final state $\psi_1^i = \psi_1$, so that $<\psi_0^i | \psi_1^i> = 0$ and $p_i(t)$ is the transition rate for such nucleus transformation. Note that in chemical reaction models, their reagents pass over potential barriers mainly due to thermal fluctuations which transfer to them kinetic energy larger than barrier height, the role of tunneling mechanism supposed to be negligible. However, they also would suffer barrier scattering, hence it's possible that similar models can be applied for them also. It's notable that in some biochemical processes quantum tunneling plays important role, example is DNA replication.

Described experimental results indicate that hypothetic NC influence perturbs parameters of system transitions, examples are chemical reactions and nucleus decays. It's reasonable to suppose that such NC influence should be reciprocal and symmetric, so that NC influence of some system $S_1$ on distant system $S_2$ evolution supposedly induced by similar transitions in $S_1$. Example of such reciprocal NC influence on tunneling rate in quantum systems was discussed in sect. 1. It's reasonable to suppose that such $S_1$ influence on arbitrary $S_2$ system would change $S_2$ transition rate, and vice versa for $S_2$ NC influence on $S_1$. We'll consider here only the simplest system transitions, i.e. from initial to final state directly without passing through intermediate stages, NC for oscillating processes will be considered elsewhere. For the start our heuristic assumption is that in that case, NC effect intensity induced by system $S_1$ will be proportional to some function of $S_1$ transition rate from its initial internal state $\psi_{in}$ to final one $\psi_f$, such that $|<\psi_{in} | \psi_f>|=0$. In particular, it can be supposed arbitrarily that NC influence intensity induced by $A_i$ nucleus decay is proportional to $p_i(t)$ of eq. (3) and it can change transition rate of some distant nucleus $A_j$. This assumption will be reconsidered below in QM formalism framework, it will be shown that it's applicable only as approximation and in general, NC influence described by corresponding QM operator. It will be argued below that observed variations of nuclear decay rates can be induced by nuclear reactions in the Sun. It's notable that some standard QM interactions are also induced by analogous transition processes, example is charged transition current in weak interactions, it stipulates, in particular, neutron β-decay (Martin, 2011). More arguments in favor of such NC mechanism discussed in (Mayburov, 2022).

Experimental results discussed in previous section indicate that NC influence can make their evolution less chaotic and more regular, in particular, it can result in squeezed states with sub-Poisson statistics. It's notable that self-ordering is quite general feature of quantum dynamics, examples are crystal lattice formation or atomic spin order in ferromagnetic. Another example is elastic particle scattering, in that case, the final state possesses higher angular symmetry than incoming plane wave state. These analogies together with cited experimental data permit to suppose that for arbitrary system evolution such NC influence tends to make its evolution most symmetric and self-ordered, our choice of NC Hamiltonians will be prompted by these assumptions. Then, one can suppose that any large system would gain via NC mechanism its own self-ordering, in particular, making its own evolution more regular and ordered. Beside the self-ordering, corresponding to more regular time intervals between events, other forms of system evolution symmetrization, in principle, can be induced by NC effects. In particular, enlargement of average time intervals between events can be also treated as the growth of temporary symmetry since event distribution becomes more homogeneous in time, its asymptotic limit is $p(t) \to const$ for $t > t_0$. It's notable that experimental results reviewed above demonstrate enlargement of nucleus life-time induced by solar activity (Fischbach, 2009; Bogachev, 2020); the same is true for its influence on some chemical reactions (Piccardi, 1962; Troshichev, 2004).

Typical experimental accuracy of nuceus decay time measurement $\Delta t$ is several nanoseconds (Martin, 2011). Formally, such measurement described as the sequence of two consequent $A_i$ state measurements divided at least by $\Delta t$ interval. If first one shows that $A_i$ is in $\psi_0$, state, and next one that it's in state $\psi_1$, it means that $A_i$ decay has occurred during this time interval (Sudbery, 1986). In QM formalism, a general state of quantum system $S$ described by density matrix $\rho$, for pure states $\rho = |\psi><\psi|$. If $A_1$, $A_2$ nuclei are its components, the partial $A_{1,2}$ density matrixes $\rho_{1,2}$ can be defined. For each $A_i$ it follows that that if other $S$ components were measured previously, then its decay probability rate would differ from $p_i$ of eq. (3) and becomes

$$\gamma_i(t) = \frac{\partial}{\partial t} Tr \, \rho_i(t) P_1^i \qquad (4)$$

where $P_1^i$ is projector on $A_i$ final state (Sudbery, 1986).

**Nonlinear QM formalism**

It was supposed that NC effects should not change the system average energy, however, if the corresponding NC Hamiltonian is linear operator then for α-decay this condition will be violated (Mayburov, 2021a). It will be shown here that nonlinear Hamiltonians can satisfy much better to this condition and so it's sensible to apply them for NC effect description. Nonlinear QM Hamiltonians were introduced initially as effective theories describing collective quantum effects, but now it's acknowledged that such hypothetical nonlinear corrections to standard QM formalism can exist also at fundamental level (Weinberg, 1989). In nonlinear QM formalism, particle evolution described by nonlinear Schroedinger equation of the form

$$i\hbar \frac{\partial \psi}{\partial t} = -\frac{\hbar^2}{2m} \nabla^2 \psi + V(\vec{r})\psi + F(\psi, \bar{\psi})\psi = H^L \psi + F(\psi, \bar{\psi})\psi \qquad (5)$$

where $m$ is particle mass, $V$ is system potential, $F$ is arbitrary functional of system state. In our case, $H^L$ is Gamow Hamiltonian with $V=U$. Currently, the most popular and elaborated nonlinear QM model is by Doebner and Goldin (DG) (Doebner and Goldin, 1992; Doebner and Goldin, 1996). In its formalism, simple variant of nonlinear term is $F = \frac{\hbar^2 \Gamma}{m} \Phi$ where

$$\Phi = \nabla^2 + \frac{|\nabla \psi|^2}{|\psi|^2} \qquad (6)$$

is nonlinear operator, $\Gamma$ is real or imaginary parameter which, in principle, can depend on time or other external factors, here only real $\Gamma$ will be exploited. Main properties of eq. (5) were studied in (Doebner and Goldin, 1992; Mayburov, 2021a), they can be summarized as follows:

(a) The probability is conserved.

(b) The equation is homogeneous.

(c) The equation is Euclidian- and time-translation invariant for $V=0$.

(d) Noninteracting particle subsystem remain uncorrelated (separation property).

(e) For $V=0$, particle plane waves are solutions both for real and imaginary $\Gamma$.

Since $\int \bar{\psi} F\psi d^3x = 0$ for arbitrary $\psi$, the energy functional for solution of eq. (5) is $<i\hbar\partial_t> = <H^L>$. Hence the average system energy would change insignificantly if not at all if $F$ added to initial Hamiltonian $H^L$, so it advocates DG ansatz application to NC models. In particular, it will be shown that in WKB approximation, which is the main formalism for decay calculus, the energy expectation value doesn't change in the presence of such nonlinear term $F$.

It's notable that nonlinear term $F$ in particle Hamiltonian can modify the particle tunneling rate through the potential barrier. In particular, analytic solution of this problem was obtained for rectangular potential barrier, in that case, the barrier transmission rate depends exponentially on $\Gamma$ (Mayburov, 2021a). To calculate corrections to Gamow model for arbitrary potential $U$, WKB approximation for nonlinear Hamiltonian of eq. (5) can be used. In this ansatz for rotation invariant $U$ α-particle wave function reduced to

$$\psi = \frac{1}{r}\exp(\frac{i\sigma}{\hbar})$$

Function $\sigma(r)$ can be decomposed in $\hbar$ order as $\sigma = \sigma_0 + \sigma_1 + ...$ , here $r$ is the distance from nucleus center (Peres 2002; Sakurai 1994). Given α-particle with average energy $E$, one can calculate the distances $R_{0,1}$ from nucleus centre at which $U(R_{0,1})=E$. Then, for our nonlinear Hamiltonian of eq. (5) the resulting equation for $\sigma_0$

$$(\frac{1}{2m} - \Lambda)(\frac{\partial \sigma_0}{\partial r})^2 = E - U(r)$$

where $\Lambda = 2m\Gamma$ for $R_0 \leq r \leq R_1$; $\Lambda=0$ for $r$ outside this interval. Its solution for $R_0 \leq r \leq R_1$ results in

$$\psi(r) = \frac{1}{r}\exp\left(\frac{i\sigma_0}{\hbar}\right) = \frac{C_r}{r}\exp[-\frac{1}{\hbar}\int_{R_0}^{r} q(u)du]$$

where $C_r$ is normalization constant,

$$q(u) = \{\frac{2m[U(u)-E]}{1-4\Gamma}\}^{\frac{1}{2}}$$

Account of higher order $\sigma$ terms practically doesn't change transmission coefficient $D$

$$D = \exp\left[-\frac{2}{\hbar}\int_{R_0}^{R_1} q(u)du\right] = \exp[-\frac{\phi}{(1-4\Gamma)^{\frac{1}{2}}}] \cong \exp[-\phi(1+2\Gamma)] \qquad (7)$$

Here $\phi$ is constant for given nucleus, whereas $\Gamma$ can depend on time and other parameters. Note that $\Lambda$ term induced by $H$ nonlinearity doesn't change particle energy in comparison with corresponding linear $H^L$. To calculate nucleus decay parameter, $D$ multiplied to the number of α-particle kicks $n_d$ into nucleus potential wall per second , it gives   $\lambda = n_d D$ (Gamow, 1928),  for DG model $n_d$ doesn't depend on  $F$ term  (Mayburov, 2021a).

## Quantum-mechanical NC model

As was noticed above, besides more regular time intervals between events, described by sub-Poisson statistics, other forms of system evolution symmetrization, in principle, can be induced by NC effects. In particular, the growth of average time intervals between events can be also considered as the enlargement of system evolution symmetry. Really, the event distribution of eq. (3) on time half-axe $\{t_0, \infty\}$ becomes maximally homogeneous for $\lambda \to 0$, hence the evolution temporary symmetry grows in this limit. It's natural to suppose that NC effect for any system of restricted or fixed size grows with the number of system constituents $N$ involved into reactions. For the case of two systems $S_{1,2}$ it's reasonable to expect that $S_1$ evolution due to NC effects would result in its own self-ordering and influences $S_2$ in the similar way and vice versa. For the case of two distant systems of which one of them $S_1$ is large and other one $S_2$ is small,  for them NC effects supposedly realized in master regime, i.e. $S_1$ can significantly influence $S_2$ state and make it evolution more ordered, whereas $S_2$ practically doesn't influence  $S_1$ state evolution. Then, resulting NC effect in $S_2$ should depend on $S_1$ evolution properties and $S_1$, $S_2$ distance $R_{12}$.

Let's study first how proposed NC influence described in master regime approximation. Consider two nuclei systems $S_1$, $S_2$ with the average distance between $S_1$, $S_2$ elements $R_{12}$, which supposedly is much larger than $S_1$, $S_2$ size. For the simplicity we'll consider here only static situations when all object positions are fixed. $S_1$ is the set of $N_1$ unstable nuclei $\{A_i\}$ prepared at $t_0$ with decay probability described by eq. (3). However, $S_1$ internal NC self-influence can change it to some $p_i^m(t)$. $S_2$ includes just one unstable nuclide B prepared also at $t_0$, its evolution described by Gamow Hamiltonian $H_b$ analogous to $H$ of eq. (2), but its decay constant $\lambda_b$ and decay probability rate $p_b(t)$ can differ from $\lambda$ and $p_i(t)$ of eq. (3). In such set-up, corresponding to master regime, NC induced by system $S_1$ supposedly influences B evolution, in particular, it can change its average life-time. Suppose that all geometric factors of such NC influence for given $S_1$ described by phenomenological real propagation function $\chi(R_{12})$ in B Hamiltonian; its absolute value supposedly reduced with $R_{12}$ and grows with $N_1$. Resulting   corrections to $H_b$ presumably are small and so can be accounted only up to first order of $\chi$. Basing on assumptions discussed above, in particular, that resulting NC effect proportional to $S_1$ transition rate and described by nonlinear B Hamiltonian term, it follows that parameter $\Gamma$ in $F$ term becomes the function $\Gamma(R_{12},t) = \chi(R_{12}) p^m_i(t)$, so that phenomenological B Hamiltonian

$$H_b^d(t) = H^L + \frac{\hbar^2}{m} \chi(R_{12}) p_i^m(t) \Phi \qquad (8)$$

where $\Phi$ is operator of eq. (6) for B. For simplicity we admit that $p_i^m(t) \approx p_i(t)$ of eq. (3) because in our model $S_1$ NC self-influence effects aren't expected to be large. Hence for such ansatz B Hamiltonian and consequently B life-time depends on $S_1$ nuclei decay probability rate. Solving Schrodinger equation with such Hamiltonian in WKB approximation for B state $\psi^b(t)$, it follows that B decay probability rate

$$p_b^f(t) = C_b \exp[-(t-t_0)\epsilon(t)] \qquad (9)$$

here $C_b$ is normalization constant

$$\epsilon(t) = \lambda_b \left(\frac{\lambda_b}{n_d}\right)^{2\Gamma(R_{12},t)}$$

Then, under $S_1$ NC  influence  resulting B nucleus life-time for $\chi>0$  becomes larger than initial one. Such $S_2$ evolution modification can be interpreted as the enlargement of $S_2$ evolution symmetry such that resulting decay probability rate  $p_b^f(t)$ would become more homogeneous in time.

Now such NC effect can be considered on more fundamental level beyond master regime. Let's suppose now that $N_1=1$ and nucleus $A_1$, B states described by wave functions $\psi^1$, $\psi^b$ correspondently. Then for the same initial conditions as above, the system initial wave function $\psi_s = \psi^1_0 \psi^b_0$ at the preparation time moment $t_0$. In accordance with our previous analysis, it's reasonable to assume that in general $A_1$ NC influence on B state evolution is proportional to $A_1$ transition rate from its initial state to final one. In QM framework, such rate described by the operator $\frac{dP_1^1}{dt}$ for projector $P_1^1$ exploited in eq. (4); it can be calculated from Erenfest theorem (Peres, 2002). Conjugal B NC influence on $A_1$ evolution has analogous dependence which defined by corresponding projector $P_1^b$ for B state. As the result, $A_1$, B Hamiltonian in first $\chi$ order

$$H_S(t) = H^L + \frac{\hbar^2}{m} \chi(R_{12}) \frac{dP_1^1}{dt} \Phi + H_1 + \frac{\hbar^2}{m} \chi(R_{12}) \frac{dP_1^b}{dt} \Phi_1 \qquad (10)$$

here $\Phi_1$ is operator of eq. (6) for $A_1$, $R_{12}$ is $A_1$, B distance; parameter $\Gamma$ substituted by $A_1$, B operators. It follows that for our system

$$<\frac{dP_1^{1,b}}{dt}> = p_{1,b}(t)$$

Since $A_1$, B operators commute, transition rate operators can be replaced by their expectation values

$$H_S(t) = H^L + \frac{\hbar^2}{m} \chi(R_{12}) p_1(t) \Phi + H_1 + \frac{\hbar^2}{m} \chi(R_{12}) p_b(t) \Phi_1 \qquad (11)$$

Solution of evolution equation for such Hamiltonian would give the probability rate for B decay described by eq. (9) with $p_i^m(t) = p_i(t)$ of eq. (3), $A_1$ decay probability rate can be calculated analogously. Note that Gamow kinematic factor $n_d$ doesn't change in this case. Thus, obtained ansatz supports applicability of eq. (5) for NC influence description for $S_1$ nuclei ensemble in master regime. It also demonstrates that our initial hypothesis that NC effect intensity related to system transition rate is reasonable, because any other option would demand more complicated Hamiltonians. Obtained $A_1$, B states are correlated but aren't entangled, so that the system state $\psi_s(t) = \psi_1(t) \psi_b(t)$ at arbitrary time, however, in the next $\chi$ order $A_1$, B state entanglement can appear. Under NC influence for $\chi>0$ resulting $A_1$, B nucleus life-times becomes larger than initial one. Such $A_1$, B evolution modification can be interpreted as the growth of their evolution symmetries such that resulting decay probabilities becomes more homogeneous in time in comparison with initial $A_1$, B decay probabilities. It can be supposed also that inverse process, i.e. $A_1$ nucleus synthesis via reaction of α-particle with remnant nucleus would induce the opposite NC effect on B, reducing B nucleus life-time and so reducing its evolution symmetry. Hence proposed NC mechanism can change, in principle, the evolution symmetry in both directions enlarging or reducing it. Thermonuclear reactions in the Sun result in production of unstable isotopes (Martin, 2011), hence according to that model, variations of such reaction intensity can result in variation of solar NC influence rate on nuclear decay parameters on the Earth. Such reaction rate variations supposedly can occur during solar flare formation, because it results in intense ejection of charged particles and γ-quanta from solar surface (Fischbach, 2009). It supposedly can be the reason for observed correlations between solar flare moments and isotope decay rate decline on the Earth (Fischbach, 2009;,Bogachev, 2020). Proposed model exploits NC dynamics based on nonlinear α -decay Hamiltonian, which, in fact, is used as effective Hamiltonian demonstrating how the decay rates can change under NC influence. It was assumed above that NC doesn't change any system average energy, in our model this condition fulfilled for WKB approximation. System momentum and orbital momentum conserved due to rotational symmetry of studied systems.

### Squeezed state production

Formal solution of QM evolution equation (1)

$$\psi(T) = W(T) \psi(t_0)$$

where integral operator of evolution

$$W(T) = \exp[-\frac{i}{\hbar} \int_{t_0}^T H(t')dt'] \qquad (12)$$

If one considers the set of *N* independent nuclei, their joint decay probability described by Poisson distribution (Martin, 2011). Consider now the case *N=2*, for the system of two independent nuclei $A_1$, $A_2$ its initial state $\psi_A = \psi_0^1 \psi_0^2$ where $\psi_0^{1,2} = \psi_0$. If their evolution is independent, then the evolution operator is the product of two right side terms of eq. (9), but it can be also formally written as

$$W(T) = C_t exp\{-\frac{i}{\hbar(T-t_0)} \iint_{t_0}^{T} [H_1(t_1) + H_2(t_2)]dt_1 dt_2\} \tag{13}$$

In our case, $H_{1,2}$ are $A_1$, $A_2$ Gamow Hamiltonians with corresponding $r_{1,2}$, $U_{1,2}$, etc. Both integrals here are from $t_0$ to *T*, the same double integral limits are used below, $C_t$ is time-ordering (chronological) operator (Feynman, 1961). In a sense, here the second integration for each term is dummy giving just multiplier $T-t_0$, yet such ansatz is used here because below NC effects will be treated via time-dependent Hamiltonians for which multiple time parameters is standard approach (Sakurai, 1994).

Let's consider just one system *S* of *N* nuclei, as was supposed, due to conjugal NC influence between its elements, its evolution can become more regular and self-ordered practically without nucleus life-time change. Because of it, *S* evolution can differ from the case of independent nuclei and would result in the temporary correlation between decays of *S* nuclei. Let's start from the simplest case *N=2* with $A_1$, $A_2$ nuclei prepared at $t_0$ at the distance $R_{12}$. Modified *S* evolution operator can be chosen from the analogy with squeezed photon state production in atomic resonance fluorescence (Mandel and Wolfe, 1995). In that case, the photon production rate is suppressed if the time interval between two consequently produced photons is less than some fixed *Δt*. Due to it, the resulting photon registration becomes more regular, and their distribution would become sub-Poissonian. Suppose that $A_1$ NC influence rate on $A_2$ characterized by some real scalar function $k(R_{12})$, its absolute value diminishes as $R_{12}$ grows, the same function describes $A_2$ NC influence on $A_1$, $k(R_{12})$ can be regarded as coordinate NC Green function. For the simplicity, we choose NC correlation of $A_1$, $A_2$ decay moments such that its evolution ansatz can be factorized into $A_1$, $A_2$ terms. For example, if no measurement of $A_1$ state is done, then resulting phenomenological $A_2$ Hamiltonian supposedly becomes

$$H_2^c(T) = H_2 + \int_{t_0}^{T} k(R_{12})\varphi(T-t')\frac{dP_1^1}{dt}\Phi_2 dt' \tag{14}$$

$\Phi_{1,2}$ is nonlinear term of eq. (6) for $A_{1,2}$, $\varphi$ is causal Green function chosen as

$$\varphi(\tau) = \eta(\tau - \upsilon) - \eta(\tau) \tag{15}$$

its possible dependence on $A_1$, $A_2$ distance neglected, assuming that it accounted by $k(R_{12})$ function. Thus, corresponding NC time dependence described as the difference of two step functions $\eta(\tau) = \{0, \tau<0; 1, \tau \geq 0\}$ which is simple variant of such ansatz (Korn, 1968). $\upsilon>0$ is NC phenomenological parameter, it corresponds to the time range in which $A_1$, $A_2$ decays are correlated; $\upsilon$ supposed to be much larger than the effective nucleus transition time (Newton, 1961). As the result, current $A_2$ Hamiltonian is time-dependent operator, so that at given time it depends on $A_1$ decay probability rate at earlier time moments. Analogous modification occurs for $A_1$ Hamiltonian with corresponding index change. Hence NC influence on $A_1$, $A_2$ Hamiltonians results in $W(T)$ modification in comparison with eq. (13)

$$W(T) = C_t exp\{-\frac{i}{\hbar(T-t_0)} \iint_{t_0}^{T} [H_1(t_1) + H_2(t_2) + (T-t_0)G(t_1,t_2)]dt_1 dt_2\} \tag{16}$$

Third term in this equality is NC term, its simple ansatz which supresses nucleus decays at small time intervals between them. Due to $A_1$, $A_2$ operator commutativity, $\frac{dP_1^{1,2}}{dt}$ operators in it can be replaced by their expectation values $\gamma_{1,2}(t)$

$$G(t_1, t_2) = \frac{\hbar^2}{m} k(R_{12})[\,\varphi(t_1 - t_2)\gamma_2(t_2)\Phi_1 + \varphi(t_2 - t_1)\gamma_1(t_1)\Phi_2]$$

Note that the second right-side term corresponds to $H_2^c$ Hamiltonian of eq. (14), $\gamma_i(t_i)$ is of eq. (4). If no measurement of $A_i$ state was performed for $t_f < T$, then $\gamma_i(t_i) = p_i(t_i)$ of eq. (3), i.e. NC influence rate is proportional to $A_i$ decay rate. Otherwise, if such measurement was done at some $t_f$ and $A_i$ was found to be in the final state, then for $t_i > t_f$ it follows that $\gamma_i(t_i) = 0$. For $t_i \leq t_f$ resulting $\gamma_i(t_i)$ follows from decay moment measurement ansatz of eq. (4) described above (Sudbery, 1986). Thus, for i=1,2 the nucleus $A_i$ NC term in first $k$ order is proportional to decay rate of neighbour nucleus $A_{2-i}$ of eq. (3). As the result, for $k>0$ the solution for joint $A_{1,2}$ decay probability rate $p_s$ will differ from independent case when $p_s(t_1,t_2) = p_1(t_1)\, p_2(t_2)$ and becomes

$$p_s(t_1, t_2) = Z\lambda^{2+2\vartheta}\exp[-g(t_1, t_2)(t_1 + t_2 - 2t_0)] \qquad (17)$$

where $Z$ is normalization constant. Under these conditions, Gamow kinematic factor $n_d$ for α-particle motion inside nucleus changes insignificantly (Gamow, 1928). Then, both for WKB approximation (Peres, 2002) and analytic solution of Gamow problem (Lubenets, 1977)

$$g(t_1, t_2) = \exp[\,(1 + \vartheta)\ln\lambda]$$

where $\lambda$ is from eq. (3) and

$$\vartheta = \frac{k(R_{12})}{2}[\eta(t_1 - t_2)\,\varphi(t_1 - t_2)\gamma_2(t_2) + \eta(t_2 - t_1)\,\varphi(t_2 - t_1)\gamma_1(t_1)] \qquad (18)$$

Due to it, if the time interval between two decay events is less than $\upsilon$, the nucleus decay rate for $k > 0$ will be suppressed and resulting decay event distribution will become more ordered. Here the second order correction of the kind $\gamma_1 \rightarrow \gamma_2 \rightarrow \gamma_1 + d\gamma_1$ are neglected. For independent nucleus decays with $N=2$ described by eq. (13) their joint decay probability corresponds to Poisson process, whereas NC dynamics term in eq. (16) would transform it to sub-Poisson one, resulting in less stochastic and more ordered event distribution. Note that resulting $A_{1,2}$ states are correlated but not entangled. For $N > 2$ the considered NC dynamics term in $W(t)$ would change to

$$G(t_1, \ldots, t_N)\, dt_1 \ldots dt_N$$

with corresponding integration over $N$ independent time parameters. As the result, for analogous $G$ ansatz the joint decay probability of arbitrary consequent decays will be suppressed for small time intervals between them, and the decay time distribution of the whole event set will be sub-Poissonian. In general case, such correlation can involve all $N$ decays resulting in more self-ordered decay moment distribution.

Two considered symmetrization mechanisms – life-time enlargement and event sub-Poisson symmetrization, in principle, can coexist and act simultaneously in some systems. Here the system self-ordering NC effect was considered, however, some distant system $S_m$ also can induce, in principle, analogous NC evolution symmetrization in system $S$, as experimental results evidence (Shnoll, 1973; Shnoll and Kolombet, 1980; Shnoll, 2009). It can be supposed that analogous NC effect description is applicable also to chemical reactions and other atomic and molecular systems. Nonlinear Hamiltonians were used here for NC effect description, however, it's possible also that in collective systems NC effects can be described by linear Hamiltonians, so that nonlinearity appears as the corresponding effective theory (Doebner and Goldin, 1996).

## Discussion

Considered experimental results and theoretical analysis evidence that novel communication mechanism between distant quantum systems can exist. It's based on new form of QM nonlocality, principally different from well-known EPR-Bohm nonlocality mechanism. In this paper, NC effects were studied for the system of metastable states, in particular, unstable nuclei ensemble, it was assumed that such mechanism can be applicable also for chemical reactions. Important feature of such NC mechanism is that such communications, in principle, can function effectively even in dense and warm media which is characteristic for biological systems. Meanwhile, multiple biochemical reactions, which can be vulnerable to described NC influence occur in them permanently. There are many confirmed effects of distant correlations in biological systems which still have no satisfactory explanation, some of them reviewed in first section. Besides, multiple publications indicate that cosmophysical effects influence also biological system development and functioning (Gallep, 2013; Hayes, 1990). Of them, it's worth to notice specially the significant influence of moon tide gravity variations $\delta g$ on seedling photon emission rate and tree stem diameter variations (Barlow, 2010; Gallep, 2013; Gallep, 2018). There is no consistent explanation now how such small gravitational force variations $\frac{\delta g}{g} \approx 10^{-7}$ can seriously affect these subtle biological processes. It's worth to notice specially that photon emission data show the essential intensity dependence on $\delta g$ time derivative (Gallep, 2018). Meanwhile, the model of nonlinear gravitational field interaction with quantum systems predicts similar gravity influence on arbitrary quantum systems, in particular, it describes in this framework the observed life-time and decay rate variations of unstable nuclear isotopes (Mayburov, 2021a). It's established now that such ultraweak photon emission stipulated mainly by biochemical reactions of protein oxidation ( Popp, 1998; Gallep, 2013). Hence it can be assumed that $\delta g$ time derivative performs analogous influence on molecular states involved in these reactions.

Another example of unsolved biophysics problems is Gurwitsch's mitogenetic effect which demonstrates that cell division in one organism or plant can stimulate cell division in other distant one separated by quartz wall (Bischof, 2003). Such wall is transparent for ultraviolet (UV) photons, their emission by living species and plants, called also ultraweak bioluminescence, was reported by many groups (Quickenden and Que Hee, 1976; Troitskii, Konev and Katibnikov, 1961) and rewieved in ( Popp, 1998). However, the detected UV photon intensity is quite low, the irradiation of organism or plant by UV lamp with similar light intensity doesn't result in any sizable effect. However, existence of NC influence between such biosystems supposedly can constitute, in fact, additional communication channel between them. First of all, it follows from considered NC properties that they can stimulate the similar self-organization processes in the neighbor detector biosystem, if such self-organization in form of cell division occurs in inductor biosystem. Suppose that in inductor biosystem the cell division occurs, so that the cell number grows. We can assume that NC self-ordering mechanism regulates it inside inductor. It's possible also that exchange of excitons with UV energy is also necessary for cell proliferation in it. Plainly, such excitons on inductor border can convert into UV or optical photons, some part of them can be absorbed by detector system. In addition, NC effect produced by inductor cells can also influence detector cells at some distance. Hence simultaneous presence of these two effects can become the reason for the cell proliferation gain inside detector system stipulated by these two different external effects. In sum, NC influence and UV radiation supposedly constitute two independent communication channels between inductor and detector.

Such quantum NC influence supposedly has universal character, in particular, such nonlocal effects can appear for systems of scattering particles. However, for metastable systems its effects are expected to be more easily accessible for practical registration due to their relatively long duration. EPR-Bohm paradox and Bell inequalities demonstrate that quantum measurement dynamics is essentially nonlocal (Bell,2004; Norsen, 2009, Gillis, 2011). It seems doubtful yet that dynamics of quantum measurements differs principally from the rest of QM dynamics, more reasonable is to expect that both of them can be described by some universal QM formalism, therefore, the presence of nonlocal terms in it would be plausible. Concerning with causality for NC communications, at the moment it's still possible to assume that such NC can spread between systems with velocity of light. But even if this spread is instant, it's notable that usually superluminal signaling in QM discussed for one bit yes/no communications (Norsen, 2009). In our case, to define the resulting change of some parameter expectation value or dispersion, one should collect significant event statistics which can demand significant time for its storing, so it makes causality violation quite doubtful possibility. In addition, NC dependence on the distance between two systems expressed by $k, \chi$ functions can be so steep that it also can suppress effective superluminal signalling. Situation can be similar to QFT formalism where some particle propagators spread beyond light cone, but due to analogous factors, it doesn't lead to superluminal signaling (Blokhintsev, 1973).